# Pathway to lowest-energy structures and stress relaxation for the surface triple junction of copper

**Authors:** Ipen Demirel,[1,†] Yuan Fang,[2,3,†] Yan Liu,[2,3,4,†] Haitian Shan,[2,3,5] Xiaopu Zhang,[1,2,3,5,*] Yuchuan Shao,[2,3] Jianda Shao,[2,3] John J. Boland[1,*]

**Affiliations:**

[1]Centre for Research on Adaptive Nanostructures and Nanodevices (CRANN), AMBER SFI Research Centre and School of Chemistry, Trinity College Dublin, Dublin 2, Ireland

[2]Laboratory of Thin Film Optics, Shanghai Institute of Optics and Fine Mechanics, Shanghai 201800, China

[3]Key Laboratory of Materials for High Power Laser, Chinese Academy of Sciences, Shanghai 201800, China

[4]School of Materials Science and Engineering, Xi'an Jiaotong University, Xi'an 710049, China

[5]Center of Materials Science and Optoelectronics Engineering, University of Chinese Academy of Sciences, Beijing 100049, China

[†] These authors contributed equally

*Correspondence to: zhangxiaopu@siom.ac.cn

**Abstract**

The behavior of surface triple junctions (STJ) at emergent grain boundaries on free surfaces is critical to the microstructure evolution, and therefore to the stability of next generation interconnects. Yet, despite this significant importance, its lowest-energy structure and local stress have remained persistently unknown. Here, we fill this critical gap through high-resolution experimental mapping of the local surface deformation at STJs of a symmetrical tilt grain boundaries with [111] rotation axis and misorientation angles 13.17°, the analysis of the local core-shifted structure and stress relaxation, and ergodic searching metastable structures accordingly with the molecular-statics (MS) calculation. The computation clearly shows the zipped Y-shaped notch as the lowest-energy structures for this group of STJs, the line profile of which is consistent with the experimental observation. This energetic preference was well explained by the stress delocalization mechanism and was tested for a wide range of boundaries with even deeper core-shift depth using machine learning methods based on the data obtained from MS calculation. By revealing the elusive thermodynamics of STJs, our findings advance the research field by redefining the energetic framework for capillary driven structure evolution and providing foundation for understanding kinetically diffusive deformation and for engineering thin-film interconnects and related materials.

## 1. Introduction

Crystal defects such as grain boundaries (GB) and surfaces are critically important because they act as preferential sites for a wide range of processes that control the material properties, stability, and performance [1-3]. Surface triple junction (STJ) or emergent grain boundary (eGB) at the free surface is the intersection between GB and the surface, alternatively termination of GB at surface. Altogether, they provide short-circuit paths for rapid atomic transport and facilitate creep deformation; leading to void formation, cavities and intergranular creep fracture because of the local stress concentration, in addition to initiating preferential chemical reactions and intergranular corrosion. For example, in nanocrystalline interconnects in advanced integrated circuits, GB and its intersection with the hetero-interfaces play pivotal roles in determining electrical, thermal and mass transport properties and detrimental processes such as electromigration [4-7].

Research on low energy structures of GBs and eGBs at free surfaces, also called STJ [8, 9] and grain-boundary surface terminations [10], can provide a better understanding of GB related material properties especially for thin films of significantly technical importance [1, 11]. In recent years, great attention was paid to determine the stable and metastable structures of GBs, GB complexions and eGBs, through experimental, calculation and theoretical methods [12-21]. For example, the out-of-plane grain rotation in nanocrystalline copper films was observed and the GB energy driven core-shift underlying this grain rotation was identified, which could play a significant role in the GB engineering and atomic scale manufacturing for the next generation integrated circuits [12, 13]. Grain-boundary phase transformation in elemental metal copper was also found, which suggests the possibility for material design [22-29]. The influence of metastable atomic structures on GB mechanical properties and diffusion clearly demonstrates the importance of the GB atomic structures [30, 31].

Interest has also been stimulated by recent experimental observations of local grain rotation, the restructuring of emergent grain boundaries and the resultant levels of stress generation, and computationally by the introduction of the core-shift method for building wedge disclination structures at STJ and the establishment of relationships between atomic calculation and the continuum elastic theory. Zhang et al. observed the local out-of-plane grain rotation at certain STJs with scanning probe microscope [14] and proposed a core-shift method to build and evaluate energetically restructured GBs at the free surface and the local stress generation [32, 33]. Boundaries with [112] rotation axes have the lowest boundary energy among boundaries obtained from the core-shift process in the $(1\bar{1}0)$ plane for (111) texture films [34]. Ignoring the transition region where the core changes direction, the core lies along the [112] direction in the close-packed plane close to the surface but lies along [111] direction in the bulk far below the surface. The [112] eGB corresponds to a zipped chevron V-shaped notch or a wedge disclination in the continuum model and hence it can be viewed as a special kind of crack-like GB diffusion wedge in the absence of external stress [35-37] or an intrinsic crack-like GB diffusion wedge driven by the variation of the boundary structure and energy. The established quadratic relationship between the eGB energy and core shift (CS) depth [32] reconciled the atomic simulation results and the continuum elastic theory, and at the same time confirmed the mechanism of the energy balance between the boundary energy decrease and the elastic energy generation for [112] core-shifted boundaries [14].

Despite this progress, several problems have yet to be solved. First, the basic idea of the balance between the elastic energy and the interfacial energy, assumed in our previous paper [14], has not been achieved and demonstrated. Atomically, the boundary core of an eGBs can lie between [111] and [112]. It need not necessarily lie along the [112] direction in the close-packed plane, particularly at the transition region, as reported previously [14, 32], and could even become curly. From the perspective of the continuum model, the STJ structure is not necessarily chevron V-shaped and thus it could have complex stress field. Secondly, there likely exists an unstable singular stress concentration at the tip of the chevron shaped wedge, as proposed by Gao et al. for unconstrained films under tensile stress [35-37]. Finally, the lowest energy STJ structures have yet to be identified.

In this paper, we first experimentally observed the local out-of-plane rotation for one small angle GB (13.17°) in a macroscopic copper bicrystal, where the global out-of-plane rotation is restricted. We analyzed the local virial stress at STJ, then revealed the stress localization at the wedge tips at an atomic level, and in the final step we lowered the eGB energy through blunting the wedge tip and delocalizing the stress field. We found that for a series of zipped Y-shaped notches derivating from the same chevron shaped notch, the eGB energy decays monotonically with the core-shifted depth.

At the same time, employing the core-shift method and instead of keeping the eGB axis along [112], we generated all possible metastable eGBs at a fixed core-shift depth to search for the lowest energy structures. For a given fixed core-shift depth we calculated, the eGB structure with the lowest energy is always coincident with the zipped Y-shaped notch. Since the number for all the possible core-shifted boundaries increases exponentially as a function of core-shift depth, to test this phenomenological hypothesis, we took the data obtained from MS calculation and trained a machine-learning model with these eGB structures described by the core-shift sequence as inputs and their energies as outputs. We tested the speculation above for STJs with three deeper core-shift depths.

This behavior indicates a fundamental tendency for this group of copper STJs of symmetrical tilt GBs with [111] tilt axis to restructure into zipped Y-shaped notches deep into the thickness of the film rather than approximated and assumed shallow zipped V-shaped notches or wedge disclinations. Given this, we expect our observations are critically important to understand the plastic behaviour and atomic fracture mechanics of thin copper films and to predict the fracture behaviour at the atomic scale [35, 38].

## 2. Method

### 2.1 Grain boundary calculation

We focused on a symmetrical tilt GB with tilt axis [111] and misorientation angle 13.17° or $2\arctan(1/5\sqrt{3})$ in this paper, since there is only one dislocation core in each periodic unit and hence the core-shift process is much simpler. The period vectors in the black and white crystals are $1/2\ [\bar{3}\bar{2}5]$ and $1/2\ [\bar{2}\bar{3}5]$, respectively. The boundary normal vectors in the black and white crystals are $1/2\ [7\bar{8}1]$ and $1/2\ [8\bar{7}\bar{1}]$, respectively, which implies the mean boundary plane is $(1\bar{1}0)$.

The molecular statics (MS) simulation were carried out with the LAMMPS software [39] and an embedded-atom-method potential for the copper developed by Mishin et al. and widely used for the defect properties of copper [40]. The atomic stress was calculated according to the virial theorem [41]. The calculated atomic structures are visualized with OVITO software [42].

GBs are calculated with periodic boundary conditions in three directions, the rotation axis, period vector and boundary normal, and each repeat cell contains a pair of parallel GBs of equal and opposite misorientation [43]. The relaxed configuration for the copper GBs is obtained by a conjugated-gradient energy minimization with respect to both the atomic coordinates and the simulation cell size along the boundary normal direction. The boundary structure searches were done with hundreds of different in-plane displacements as the initial configuration. Only the lowest-energy structures are considered to build further the suspended films for eGB research. After the relaxation of the boundary with three atomic layers along the tilt axis [111], the bottom layer is labelled as A and the top one C. We built the suspended film through stacking the relaxed three-atomic-layers together and then adding the vacuum layer. More detailed information could be found in our previous work elsewhere [32].

### 2.2 Emergent grain boundaries and core-shift method

The core shift method was used in this paper, as shown in Fig. 1. Figure 1a shows the out-of-plane grain rotation from the black bicrystal to the red bicrystal and consequent GB core shift with an angle $\sim\psi$. The tilt axis for the black bicrystal is [111] and the misorientation angle $\theta$. The two black half-bicrystal were rotated along the x axis $\varphi/2$, clockwise and anti-clockwise, respectively. The tilt axis shifts an angle $\psi$ from [111] towards [112] or even further in the $(1\bar{1}0)$ plane. The core shift process in a single atomic plane perpendicular to the tilt axis can be achieved by removing (see Fig. 1b) or adding atoms. Clearly the dislocation line movement direction is perpendicular to the Burgers vector and hence the core-shift process is a special kind of dislocation climb. The shifted GB core can be clearly seen in the GB plane, as shown in Fig. 1c. Figure 1d shows a perspective view of the core-shifted emergent boundary for GB [111] 13.17°. The step-by-step process to build a [112] CS boundary from the suspended film with a pair of [111] boundaries is shown in Fig. 1e. The top, middle and bottom panels are projected along the tilt axis, the boundary normal and the period vector, respectively. Below each three-view visualization the corresponding cross-sectional view is shown. The atoms to be deleted in each step were located between the two red arrows marked in the graph.

The defect energy or the excess energy as a result of the presence of the interfaces and the four ideal eGBs is $E_N$, where $N$ is the total number of atoms in the system. We also calculated the defect energy $E_{N-D}$ of suspended films, which includes a core-shifted eGB with $D$ atoms deleted. The defect energy variation

$$\Delta E \triangleq E_{N-D} - E_N = (\epsilon_{cse} - \epsilon_{ie}) \cdot l \quad (1)$$

is the extensive energy difference between the core-shifted eGB energy $\epsilon_{cse}$ and the corresponding relaxed ideal eGB energy $\epsilon_{ie}$, where $l$ is the triple junction length in the calculation.

The labelling system details both the depth of the core shift and the number of deleted atoms. For example, the core shift depth (CSd) for the second structure in Fig. 1e is the top 3 layers, the number of the deleted atoms (DA) is 3, and the stacking fault depth is 6 atomic layers. We denote this structure as CSd3DA3. To build CSd6DA9 from CSd3DA3, 6 atoms between two red arrows in the middle panel of the cross-section view have to be deleted. The perspective view of the structure CSd9DA18 is shown in Fig. 1d. Atomically, the number of deleted atoms each layer in the first ABC stacking unit of CSd3DA3 is 1, which also means the core in the first ABC stacking unit was shifted one time. The number of deleted atoms each layer in the first ABC stacking unit for CSd6DA9 is 2, and that in the second ABC stacking unit is 1. This means that the core in the first ABC stacking unit was shifted two times and that in the second ABC stacking unit was shifted one time. Similarly, the number of deleted atoms each layer in the first, second, third ABC stacking unit for CSd9DA18 is 3, 2, 1, respectively, as shown in the middle panel. We have also added a V-shaped notch on the right side of the atom deletion pattern or the core-shift sequence. The numbers of deleted atoms each layer in the first, second, third, fourth ABC stacking unit for CSd12DA30 are 4, 3, 2, 1, respectively.

### 2.3 Machine learning method

Machine learning techniques are widely used in research areas such as material design [44-46] and GBs [47-51]. Here, we used scikit-learn Python library to implement the machine learning algorithm [52]. The machine learning implementation flowchart is shown in Fig. 2. It contained six main steps, these are: data collection, feature selection and descriptor, algorithm, training and evaluation, machine learning model, and prediction. Our enumerated eGB structures and their energy obtained from the MS calculation form the dataset (see Tab. S2-3). We chose the core shift times or number of deleted atoms for each ABC stacking unit as the key features and the sequence of the core-shift times from top surface as the descriptor. Different algorithms were employed such as neural network of different layers, random forest and least square regressions (LSR) of different orders, as shown in Fig. S1-2 in Supplementary Materials (SM). For the machine learning algorithm of LSR, the explicit expression for the model was also exploited. For model training, the whole data sets are divided into training sets and test sets by random sampling. The training data sets contained 80% of all the data and the rest belong to the test sets.

## 3. Results and discussion

### 3.1 Experiment

Considering an eGB structure at the fixed depth in the continuum theory, as shown in Fig. 1a, the out-of-plane rotation with a certain angle shifts the combined rotation axis from [111] toward [112] in the $(1\bar{1}0)$ plane. The eGB defect energy is the balance of the GB energy and resulting elastic energy consequently generated. For the [112] core shifted eGBs, the GB energy reaches the lowest value [34]. However, in this case, the elastic energy $E_{el}$

$$E_{el} = G/(4\pi(1-\nu)) \cdot \varphi^2 h^2 \tag{2}$$

for the wedge disclination corresponding to the zipped chevron V-shaped notch is generated, where $G$ is the shear modulus, $\nu$ the Poisson ratio, $\varphi$ the out-of-plane rotation angle of the core shifted eGB, and $h$ the wedge depth [32]. Since the GB energy decreases monotonically as the out-of-plane rotation angle increases towards $\varphi$, there is a need to balance the boundary energy and the elastic energy with an out-of-plane rotation angle between 0 and $\varphi$. While in our previous papers [14, 32, 34] and one pioneering paper about mechanical behavior of STJs [53], it is approximated and assumed that the STJ structure is [112] core shifted eGB.

Since there are four non-equivalent dislocation cores in each period for the emergent boundary with [111] tilt axis and the 26.01° misorientation angle previously reported [14], it is too complex to be taken as an example to show the energy balance scheme [14]. Hence, we designed a macroscopic bicrystal sample with [111] tilt axis, 13.17° misorientation, 8 mm diameter and 2 mm thick, as shown in Fig. 3a [32]. The experimental method employed for GB [111] 13.17° is the same as that for GB [111] 26.01°, as described before [14]. An engineered copper bicrystal sample was prepared by MaTeck GmbH according to our design specification. The sample is prepared with cycles of Ar+ ion sputtering and annealing up to ~ 800 K. The data presented in Fig. 3b are collected with scanning tunnelling microscopy at 77 K under scanning conditions 20 pA and -0.01 V. The noise level along surface normal is less than 5 pm. The surface was atomically clean without surface and subsurface impurities [14]. There is only one dislocation core in each period for this emergent boundary or five close-packed atomic lines, as analysed in our previous paper and hence this bicrystal is critical for the low-energy STJ structure problem [32]. Experimentally, the eGB also exhibits local grain rotation from the perspective view and line profiles, as shown in Fig. 3b-c. While the local grain rotation is evident, the dissociation width is close to the local rotation width, which is a little different from that for the surface profile of GB [111] 26.01°. Hence, we take GB [111] 13.17° as an exemplar to search the lowest-energy structure.

3.2 Cell size effect and local stress analysis

Before the local stress analysis and searching for the lowest-energy structure, we first carried out simulations with the core-shift method and calculated the eGB energy with different cell sizes along the boundary normal direction, to remove any artificial effects due to the boundary interaction between two GBs in each simulation cell. The cell sizes along the boundary normal are 3, 5, 10, 15, and 20 times the magnitude of the boundary normal. The quadratic term of the parabolic curves involves the elastic energy and it is influenced by the boundary interaction, as shown in Fig. 4a. Through parabolic fitting, we obtained the coefficient for the quadratic term. It is clear that there is boundary interaction when the cell size along the boundary normal is less than 10 times the boundary normal magnitude. Otherwise, the coefficient saturated to the value ~0.041, consistent with our previous calculation, and hence the effect of boundary interaction can be neglected. The relationship between the core-shift depth ($CSd$) and the wedge depth $h$ for the continuum disclination model

$$h = (CSd + 1.5) \cdot a/\sqrt{3} \tag{3}$$

was established in the previous paper and are plotted in the upper and lower horizonal $x$ axes in Fig. 4 [32]. In the simulations below, we took the simulation cell size along the boundary normal to be 15 times the boundary normal magnitude.

The relaxation effect of the simulation cell along the boundary normal is also shown in Fig. 4b. The black curve and the red curve show the eGB energy without and with the simulation cell relaxation, respectively. The relaxation reduces each eGB energy slightly. The thickness effect was also examined, taking GB [111] 9.43° for example. We calculated core-shift depth dependent eGB energy at three different film thickness, that is 80, 120 and 160 ABC-stacking units, as shown in Fig. 4c. Clearly, another free surface helps the system to relax further. To avoid this effect, we used the film of a sufficient thickness with 160 ABC-stacking units. In summary, the size effects are removed by increasing the thickness of the suspended films and increasing the spacing between two equal and opposite boundaries in each simulation cell.

We firstly performed a local stress analysis in order to reduce the local elastic energy. A wedge disclination with tilt axis [112] near the surface and tilt axis [111] deep into the film was built previously and corresponds to a perfect chevron V-shaped notch in the continuum model. We immediately noticed the similarity to the crack-like GB diffusion wedges subjected to external stress for the eGB in metal films proposed almost three decades ago by Gao [35] and other related models [54, 55]. The analysis there shows a singular stress concentration near the wedge tip that is similar to the case in a crack, but shows a stress concentration at the wedge opening. In contrast, a real crack opening is nearly stress-free [37, 56, 57].

Considering the core-shift method, these low energy [112] core-shifted GBs, and their relationship with the continuum wedge disclination model, we analysed the hydrostatic stress field at different core shift depths, as shown in Fig. 5, during the low energy structure search. All images were colored with atomic hydrostatic stress from -0.5 GPa to 0.5 GPa. The blue atoms are under compressive stress and the red atoms are under tensile stress. The core shift depth is followed by the total number of deleted atoms. The atomic structures for CSd0DA0, CSd3DA3, CSd6DA9, CSd9DA18, CSd12DA30 were shown in Fig. 1e and the corresponding eGB energy was shown in Fig. 4b. The white arrows in Fig. 5 identified the deepest core-shifted layer. Clearly, the hydrostatic stress field transitioned from tension above the arrow to compression below the arrow. It makes sense that the upper closed or zipped region is under tension and the tip of the chevron V-shaped notch is under compression in continuum mechanics due to the shear deformation involved during zipping up the chevron V-shaped notch [14, 58]. From these images, it is clear the compressive stress at the wedge tip is much more localized than the tensile stress at either side of the wedge opening.

We found similar stress concentrations at the wedge tips for all [112] core-shift eGBs. Since the elastic energy depends quadratically on the stress, it lowers the system energy to delocalize the stress. In the elastic theory of dislocations, the wedge or crack can be viewed as a short array of edge dislocations or discontinuous tilt boundary with both Burges vectors and dislocation senses parallel to the surface plane, and Burges vectors perpendicular to the dislocation senses. In a discontinuous tilt boundary ending inside the crystal, the pile-up produces a large normal stress at the tip, which results in a strong climb force on the leading dislocation, tending to force the leading dislocation away from the wall and leave a dislocation free zone [59, 60].

### 3.3 Zipped Y-shaped notches and stress relaxation

To fulfill the above analysis, we delocalized the stress concentration by blunting the wedge tip by transforming it into a zipped Y-shaped notch. The blunting process for a [112] core-shifted eGB derived from GB [111] 13.17° is shown in Fig. 6a. We took as an example the structure CSd12DA30, which is [112] core-shifted emergent boundary. The numbers of deleted atoms each layer in the first, second, third, fourth, and fifth ABC stacking unit for CSd15DA33 are 4, 3, 2, 1, and 1, respectively. The numbers of deleted atoms each layer in the first six ABC stacking units for CSd18DA36 are 4, 3, 2, 1, 1, and 1, respectively. The remaining eGBs in the list can be analyzed similarly.

If we consider the wedge disclination for [112] core-shifted eGBs as the zipped chevron V-shaped notch CSd12DA30 shown in first panel in Fig 6a, the remaining panels show the progressive formation of zipped Y-shaped notch structures. The core-shifted eGB energy as a function of the core-shift depth for the zipped Y-shaped notch derivating from CSd12DA30 was shown as the blue curve in Fig. 6b. It decreases monotonically. The longer the vertical stem of the Y-shaped notch, the lower the corresponding eGB energy, shown as successive blue points along the blue curve. Note that the [112] core-shifted eGB energy itself is shown as the black parabola-like curve in Fig. 6b. We also carried out computation on the eGB energy of the Y-shaped notch structures derivating from other [112] core-shifted eGBs, and similar behavior was found (see Fig. 6b).

We always observed a monotonic change in energy while increasing the vertical stem length of the zipped Y-shaped notch regardless of whether or not we started with the lowest energy zipped V-shaped notch (CSd=12). In some cases, for example CSd=3 and 6 the red and green curves in Fig. 6b the energy increased monotonically while for CSd =9, 12, 15 and 18 (orange, blue, cyan and purple in order), it decreased monotonically. The transition depth is between 6 layers and 9 layers. We found that all the curves can be well fitted to an exponential formula (See Tab. S1).

In addition to the emergent boundary eGB [111] 13.17°, we also carried out simulation on eGB [111] 9.43° and eGB [111] 3.89°. They exhibit the same behaviour for the [112] core-shift boundaries, as shown in Fig. 7(a-b). The group of zipped Y-shaped notches derivating from the same [112] core shifted boundary also show exponential relation between the eGB energy and core-shift depth. The transition depth highlighted by the arrows for eGB [111] 3.89° and eGB [111] 9.43° are a value between 9 and 12, and 9, respectively. Comparing the three emergent boundaries, the transition depth hardly depends on the misorientation. In addition, the transition depth is clearly different from the vortex depth with lowest eGB energy on the parabolic curves. Both suggest the transition depth is a result of surface effects.

To find the principle underlying for the monotonic energy decrease of the eGB structure shown in the curves in Fig. 6b, the zipped Y-shaped notch with CSd=12 was scrutinized. Note that the line profiles of CSd12DA30, CSd15DA45, and their corresponding zipped Y-shaped notches closely track those from the experimental counterpart as shown in Fig. S4, despite a simplified simulation does not account for the influence of temperature, atomic potentials, or external stress on the depth and width of the grooves [40, 53]. In another word, the experimental line profile cannot uniquely determine a single metastable zipped Y-shaped notch. We calculated the atomic scale stress for three structures,

CSd12DA30, CSd18DA36, CSd24DA42, corresponding to different points along the blue curve in Fig. 6b. The results are shown in Fig. 6c. It is clear that when the stem of the Y-shaped notch becomes longer, both the stress intensity and the area are reduced significantly, in particular the compressive stress intensity and its corresponding area. For example, the numbers of deleted atoms each layer in the first few ABC stacking units from the top surface for CSd12DA30 are 4, 3, 2, and 1, respectively. The numbers of deleted atoms each layer in the corresponding ABC stacking unit from the top surface for CSd18DA36 are 4, 3, 2, 1, 1 and 1, respectively, and for CSd18DA42 they are 4, 3, 2, 1, 1, 1, 1 and 1, respectively.

Our results above show clearly that in most cases the wedge disclination corresponding to the zipped chevron V-shaped notch with the depth greater than the transition depth is not the lowest energy structure. Instead, the eGB defect restructures into a zipped Y-shaped notch with an even lower defect energy. If we consider the zipped Y-shaped notch as a new grain boundary phase [61], the total GB energy actually increases, since the core-shifted grain boundaries with [112] rotation axis is the energetically preferred. The transformation from the zipped V-shaped notch to the zipped Y-shaped notch actually increase the total boundary energy but reduces the elastic energy and the total defect energy. However, the lowest eGB structures for a certain core shift depth are still unknown, considering the large numbers of metastable structures.

### 3.4 Energy balance, enumeration and speculation

Instead of searching for the lower energy eGB structures through delocalizing the stress field in the atomic calculation, an alternative approach involved fixing the core-shift depth and enumerating all the possible core-shift combinations to find the stable eGB structures. This method works well, in particular for shallow core shift depths. The results are shown in Fig. 8. Figure 8a shows all the possible eGB structure from the lowest energy toward the highest energy, obtained with the core-shift method at a fixed core shift depth CSd = 12, which means there are no atoms deleted from layers deeper than this. Note that two eGB structures, CSd12DA18 and CSd12DA27 (circled in red in Fig. 8b), share nearly the same eGB energy even with different numbers of jogs in the eGB stacking fault ribbon. The less point defects along the eGB stacking fault ribbon generally correspond to the eGB close to [112] core shifted one and less elastic energy. Evenly distributed point defects rather than the agglomerated point defects along the eGB stacking fault ribbon generally lower the system elastic energy. Since the elastic energy is not only dependent on the number of jogs but also their relative position and the system energy includes both the GB energy and the elastic energy [59, 62], there is no simple way to estimate the total eGB energy from the structures in Fig. 8a.

The numbers of eGB structures in total for a fixed core shift depths of 3, 6, 9, 12, 15, 18, 21 layers are 1, 2, 4, 8, 16, 32, and 64, respectively. We did not carry out ergodic simulations for lowest energy eGB structures for a fixed core shift depth greater than 21 layers due to the exponentially increasing number of structures and consequently the huge amount of computation work required. To ensure that we have listed all the possible metastable core shift structures, we developed an enumeration rule. Our calculation shows that metastable core shift structures can be obtained only when the core-shift times for the ABC stacking unit above is the same as that below or 1 more than that below. Accordingly, the total numbers of eGB structure at certain core shift depth is

$$N(CSd) = 2^{(CSd/3)-1} \quad (4)$$

For example, the number of eGB structures for a fixed core shift depth 21 is $N(21) = 64$, which is the same as that in the enumeration method. It indicated that we indeed have listed all the possible core-shifted structures. Furthermore, we also tested eGB structures violating the enumeration rule and found that eGB energies are always increased and varies greatly.

The enumeration approach shows the coincidence of the lowest eGB energy at certain core-shift depth with the asymptote of the exponential curve describing the zipped Y-shaped notches deriving from each zipped V-shaped notch, such as that described in Fig 8b. We further anticipated that the lowest energy structures, at depth greater than the vertex depth involving the bottom of the parabolic curves for the [112] CS boundary, correspond to the data point along the exponential line. Apart from the above analysis, we noticed that other clear patterns emerged in Fig. 8b. For example, the exponential relationship shown in Fig. S3 and by further examination we found this group of structures shared the same notch shape.

We summarized our research graphically for an eGB with the local tensile stress, as shown in Fig. 8c. There is a natural transition of the stable eGB structure from the zipped V-shaped notch to zipped Y-shaped notch. For either the V-shaped notch or the Y-shaped notch, the continuum model includes four steps: these are the initial structure, cut-out, zipping and final structure with local tensile stress, which is similar to the published method [14, 53]. The zipped V-shaped notch can be represented by a regularly spaced wall of edge dislocations and the zipped Y-shaped notch by a relaxed wall of edge dislocations. In one extreme case, the zipped V-shaped notch dominates, as proposed by Gao, et al [35]. In other extreme case, the zipped Y-shaped notch is deep down into the bottom of the films, which is the case proposed by Chason, et al [63-65].

### 3.5 Machine learning results and testing

Considering the time cost to relax a single STJ structure and its energy, it's not feasible to work on the large number of metastable STJ structures with MS calculation. There are 64, 128, 256 and 512 metastable STJ structures for CSd 21, 24, 27, 28, respectively. To predict the energy for different eGB structures at core shift depths beyond 18 layers, we built a machine learning model as a surrogate predictor based on results obtained with MS calculation. It includes four main steps. The first step is data collection from LAMMPS calculation as we described and analysed above. The remaining three steps are detailed below.

The second step is to select the feature to represent our eGB structures. From the above analysis, it is clear that the core-shift times for each ABC stacking unit is one option. The descriptor for one eGB structure is then the sequence of its core-shift numbers, one number for each ABC stacking unit. The dimension of the descriptor is the number of ABC stacking unit involved in the core-shift process. The eGB energy is the corresponding label value for this eGB structure. For example, the descriptor or the feature vector for the second structure in Fig. 8a is a vector [4, 3, 2, 1], corresponding to a number 4321, for the training of a 4-dimensional model. The corresponding descriptor was transformed into [4, 3, 2, 1, 0, 0] for the training of a 6-dimensional model through dimensionality increment, corresponding to a number 432100.

The third step was to select the machine learning algorithm. We tried single layer neural network, random forest and different kinds of LSR such as linear, polynomial, logarithmic and exponential prototypical function. For the neural network model, we adopt one hidden layer with 70 thousand neurons after trialing with several different parameters. For random forest model, the number of trees in the forest is 50. For the regression algorithm, the best result is obtained with the polynomial function of the descriptors and the degree 2, or second order polynomial. Hence, only the results from LSR with degree 2 were shown below.

The fourth step was to train a model and evaluate the model. The performance of the model is evaluated by the root mean square error (RMSE), the determination coefficient $R^2$ and comparison with calculated values with LAMMPS.

Figure 9a shows the comparison between the calculated values and the predicted values from machine learning models with three different algorithms, as a function of the numerical order. The eGB structures were ordered according to the corresponding numbers of the descriptor vectors. There are $CSd(21) = 64$ predicted data points in total for core-shift depth 21 (see Tab. S4). The machine learning model is obtained from the training dataset for core shift depth 3, 6, 9, 12, 15, and 18. It includes all the data with CSd less than 21 and the additional 6 data points on the exponential lines with CSd=21 in Fig. 8b. The total number of data is $\sum_{i=0}^{5} 2^i + 6 = 69$. The black data shows the calculated data with MS method. It is clear that predicted data from LSR algorithm and random forest algorithm looks much better than that from the neural network.

To qualitatively address the prediction accuracy and performance of different algorithms, we plotted out the predicted eGB energy as a function of the calculated eGB energy and calculated the determination coefficient for different algorithms, as shown in Fig. 9b. The determination coefficient $R^2$ is calculated on the test data only. The solid line is the identity line and dashed lines show $\pm 0.02\ nJ\ /\ m$ away from it. From the plot, obviously the LSR algorithm underestimates the eGB energy and works quite well in all range of the eGB energy, while the random forest overestimates the eGB energy for low energy eGBs and severely underestimates the eGB energy for high energy eGBs. The data points from the neural network prediction scatters around both sides of the identity line. The RMSE for LSR, random forest, and neural network are 0.0028 $nJ/m$, 0.0134 $nJ/m$ and 0.0173 $nJ/m$, respectively. The determination coefficient for LSR, random forest, and neural network are 0.996, 0.917, and 0.860, respectively, which indicates that the LSR is the best choice.

To predict the eGB energy and test our speculation, we trained other models with all the data with core-shift depth no more than 21 with a descriptor vector dimension no more than 7. Since the high determination coefficient of the LSR of degree 2, we trained new model with this algorithm only. For example, to predict the eGB energy with dimension 8 and CSd 24, we transformed the descriptors by increasing the dimension of all the data with CSd less than 24 into 8 by add zero at the end of the descriptor vectors and add the 6 points with dimension 8 on the exponential lines. The predicted 128 results were shown in Fig. 10a. The global minimum energies are indicated by arrows, which is one of the circled data points on the exponential lines in Fig. 8b.

Similarly, we increased the descriptor vector dimension again, add 6 more available data points with CSd=27 on the exponential lines, and then trained a new model for CSd no more than 27 and dimension 9. The predicted energy for all 256 structures with CSd=27 is shown in Fig. 10b and the minimum energy is highlighted with arrow. This energy is also located on the exponential line in Fig. 8b and the structure corresponds to the zipped Y-shaped notch.

To obtain a new model for CSd no more than 30 and descriptor vector dimension 10, we add 18 data points on the exponential lines with CSd=24, CSd=27 and CSd=30 to the previous training dataset with CSd no more than 21. The predicted energy for all 512 structures with CSd=30 is shown in Fig. 10c and the minimum energy is highlighted with arrow. This energy is also located on the exponential line in Fig. 8b and the structure corresponds to the zipped Y-shaped notch as well. At last, we added all the predicted data with CSd=24, 27, 30 into the plot shown in Fig. 10d. Clearly the lowest-energy structures for STJs are all located on the blue exponential curve in Fig. 8b.

$R^2$ on test data for above three models are 0.99984, 0.99978 and 0.99976, respectively. They are extremely close to 1. RMSE on test data for above three models are all ~ 0.001, much smaller than that of STJ energy. Given these and LSR with degree 2, there possibly exists an excellent mathematical relationship. Hence, we analysed the explicit expressions with the algorithm of LSR. The coefficient for the four models obtained above are very similar, which is detailed in Tab. S5-8. The statistical analysis based on bootstrap resampling and structured validation shows the ratios of extrapolation RMSE and cross-validation RMSE above are ~ 2.1, 5.2 and 4.3 for 8-, 9-, 10-dimension models, as shown in Fig. S5.

### 3.6 Quantitative energetic analysis

Apart from the structural analysis based on experimental data and MS calculation, the qualitative energetic analysis through the MS calculation, stress analysis and dislocation theory [14, 32], the quantitative energetic analysis, such as comparing that calculated with MS and finite element method (FEM) is still missing.

To calculate the elastic energy of STJs for all the Y-shaped notches with MS, we first calculated the total energy of STJs for both zipped V-shaped notches and zipped Y-shaped notches. Secondly, we fit the depth dependent STJ energies for the zipped V-shaped notches with the parabola equation to separate the total STJ energies into the quadratic term, linear term and constant, which correspond to the elastic energy, interfacial energy and linear defect energy [14, 32]. Finally, we found that the interfacial energy of the stem part of the zipped Y-shaped notch was the same as that in bulk, and assuming the linear defect energy hardly change during the transformation from the zipped V-shaped notches to the zipped Y-shaped notches, the elastic energy for the zipped Y-shaped notch was obtained by removing the interfacial energy and linear defect energies for their corresponding zipped V-shaped notches.

To calculate the elastic energy of STJs for all the Y-shaped notches with FEM, we employed the similar method in our previous paper [14]. The elastic field near the valley is calculated using the linear elastic constitutive law in the finite element method (FEM) package ABAQUS. The bicrystal film is modelled using 4-node bilinear plane strain quadrilateral elements.

The results are shown in Fig. 11. The elastic energies for different STJs calculated with FEM are compared with that calculated with MS, as shown in Fig. 11. Results from both methods show quantitative agreement numerically and similar trend as the core shift depth increases during the transformation from the V-shaped notch to the Y-shaped notch. It indicates that elastic-energy results from MS are consistent with that from continuum mechanics.

### 3.7 Discussion

The kinetics of this GB rotation with the rotation axes perpendicular to the tilt axes as shown in Fig. 1a is probably achieved through collective atomic diffusion along the dislocation line and the consequent dislocation-line movement perpendicular to the Burgers vector or dislocation climb [13, 35, 36, 53, 56, 66, 67]. In this process, the local rotation axis varies and the misorientation angle changes as well. In contrast, during grain rotation under an external stress field, as observed by Wang et al., the rotation axis remains the same while the misorientation angle varies [66, 68, 69]. Furthermore, the dislocation climb direction proposed by Gao et al. is along the film normal. Therefore, this climbing direction is perpendicular to that in the core-shift process and both are perpendicular to the Burgers vector [35-37].

Diffusive processes associated with atoms entering and exiting the GBs and the transformation from zipped V-shaped notches into zipped Y-shaped notches are intrinsically linked to the GB transformation [16, 22, 24, 31, 70, 71]. Under different constraints such as thermal treatment, external force field and atom flux during deposition, the free energy of eGB structure varies and hence eGB keep restructuring themself through eGB transformation and boundary transformation.

It is reported that when diffusional deformation processes such as GB diffusion or surface diffusion dominates over dislocation plasticity, a crack-like GB diffusion wedge forms under the tensile force along the boundary normal [35]. The wedge generates a stress field with crack-like singular stress concentration at its tip in the continuum theory. This stress field promotes the dislocation nucleation at the root of the grain boundary. In contrast, our finding shows that when the stem of a zipped Y-shaped notch rather than the tip of the zipped V-shaped notch reaches the interface between the film and the supporting substrate, the structure decomposed into a single parallel-glide dislocation and a new zipped Y-shaped notch [35]. Then the parallel gliding of the parallel-glide dislocation takes place [35, 38]. The necessary stress for the parallel-glide dislocation nucleation is expected to decrease greatly.

The discovery of the intrinsic crack-like GB diffusion wedge in the absence of external stress field highlights the energy transformation between defect structures and local stress distribution and may prompt a re-evaluation of the tensile or compressive stress generation [54, 63, 64, 72, 73] and even the Griffith fracture theory at the atomic scale. This is particularly relevant for the scenarios where the eGBs can survive, such as in nanoscale devices and the additive manufacturing [74].

STJs are a fundamental type of defect on any real workpiece made of polycrystalline materials, and they are especially prevalent in thin films. They play a significant role in the local stress and microstructure evolution. Our research has possible implications for understanding and controlling a wide range of surface phenomena such as thermal grooving, interfacial corrosion, stress corrosion, interfacial fracture, surface catalysis, interconnect failure, and even creep fracture in the engine. Thus, this work is valuable not only to material science and engineering but also to broader engineering design practice. Finally, further research on STJs and our proposed model is highly warranted, such as the effect of impurities and alloy elements on electromigration, particularly manganese and cobalt, under service in advanced interconnects [75].

In our experiment, the local grain rotation at STJ for eGB [111] 13.17° was observed, which implied a core-shift STJ and the structural relaxation. Based on the observed line profile of local deformation and local stress analysis, our atomic simulation shows that lowest-energy STJ structures corresponds to the zipped Y-shaped notch for symmetrical tilt grain boundaries with [111] rotation axis and misorientation angles less than ~32.1°, although it is not uniquely confirmed experimentally. Massive calculation with MS method showed that there is a preference for low-energy core-shifted GBs in the bulk and at the (111) surface of fcc metals with misorientation angles less than 32.21°. Considering the scaling effect of the grain boundary energy for fcc metals [34, 76, 77] and given the analogous stress concentration and relaxation behavior for the zipped V-shaped notch, we propose a working hypothesis that zipped Y-shaped notches possibly constitute the lowest-energy STJ configurations for other fcc metals, which remains an open question that warrants further systematic investigation, such as MS calculation with empirical potential [78, 79].

## 4. Conclusion

In conclusion, we observed the local deformation at STJ experimentally and analyzed the boundary core-shifting and stress relaxation. The corresponding MS computation, combined with experimental observation, shows that the STJ adopts a zipped Y-shaped configuration for symmetrical tilt grain boundaries with [111] rotation axis and misorientation angles less than ~ 32.21°, which could rewrite how we model stress relaxation and failure in nanointerconnects. By searching the metastable eGB structures computationally, we found the stable eGB structures from all possible metastable structures at various core-shift depths and identified the zipped Y-shaped notch as the stable eGB structure for this group of STJs, the line profile of which is consistent with, although not uniquely confirmed by, the experimental observation. Through machine learning method based on the MS-calculated data, we tested the above speculation of STJs with even deeper core-shift depth. We also revealed that the delocalization of the stress concentration promotes the transformation of the eGB structure from the zipped V-shaped notch to the zipped Y-shaped notch, which strongly supports the presence of restructured emergent boundaries corresponding to a zipped Y-shaped notch in (111) texture films subjected to external fields.

Our research helps understand the surface-assisted GB phase transformation, stress localization and failure in thin films. It also helps analyze kinetic processes, such as dislocation climb and GB diffusion, and related local stress relaxation. Finally, it demonstrates via calculations and visualization for the widespread metastability of emergent boundaries in thin copper films and stimulates the strategies for related material engineering [25, 28, 29, 80-85].

**CRediT Author Statement**

**Ipen Demirel:** Investigation, Software. **Yuan Fang:** Investigation, Software. **Yan Liu:** Investigation, Software. **Haitian Shan:** Investigation, Software. **Xiaopu Zhang:** Funding acquisition, Supervision, Writing. **Yuchuan Shao:** Funding acquisition. **Jianda Shao:** Funding acquisition. **John J. Boland:** Funding acquisition, Supervision.

**Declaration of competing interest**

The authors declare that they have no known competing financial interests or personal relationships that could have appeared to influence the work reported in this paper.

**Data availability statement**

Data will be made available on reasonable request.

**Acknowledgements**

X.Z. and Y.F. acknowledge the funding support from National Natural Science Foundation of China (Grant No. 52573361). X.Z. are grateful to Jian Han from City University of Hong Kong and Jinbo Yang from Ningbo Tech University for helpful discussion. J.J.B and I. D. acknowledge the support from Science Foundation Ireland (Grants No. 12/RC/2278_2).

**Figure 1 core-shift method to build emergent grain boundaries**

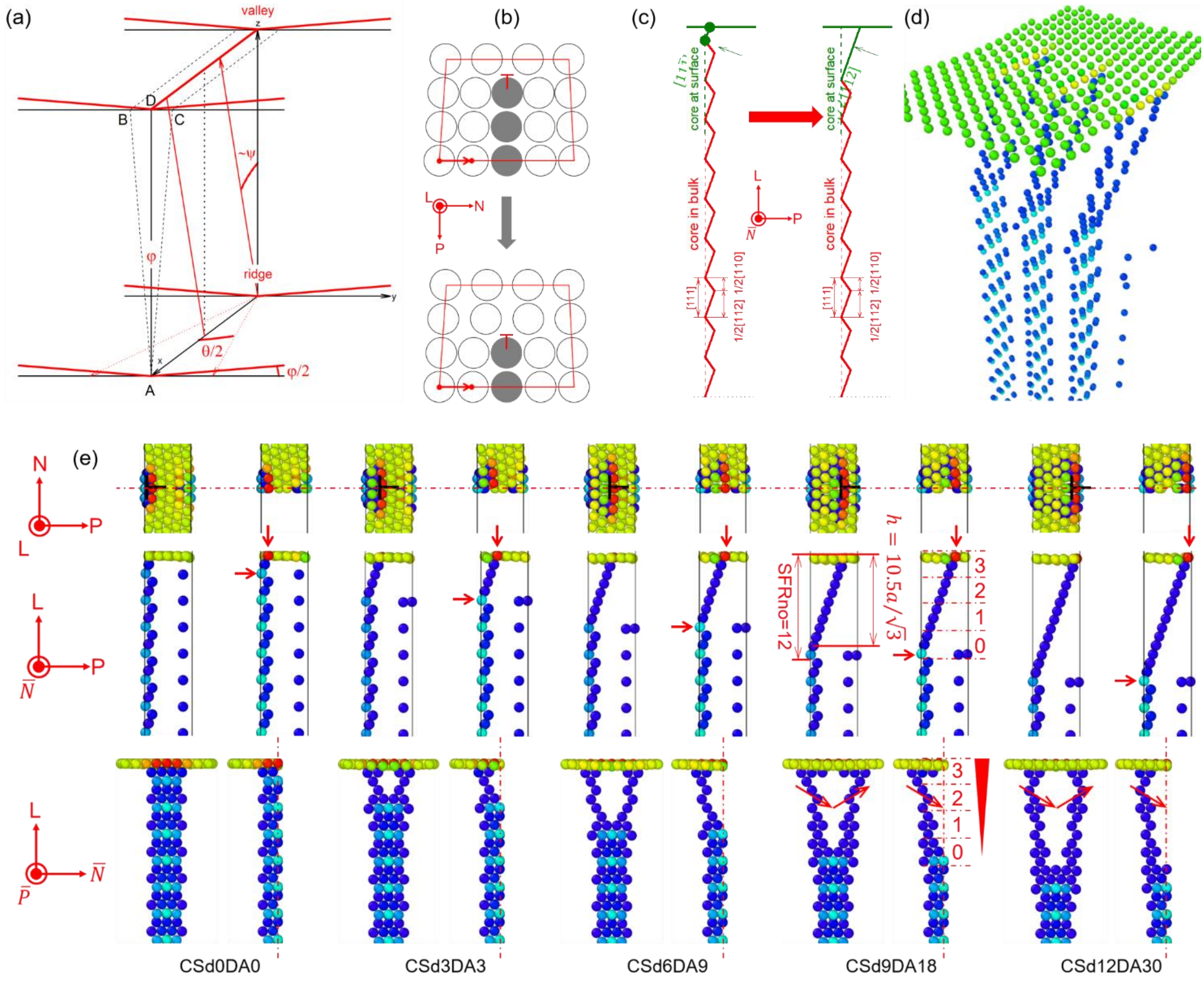

(a) out-of-plane rotation along the x axis of a bicrystal with [111] axis and misorientation angle $\theta$, and consequent core shift from [111] toward [112] and even further in the $(1\bar{1}0)$ plane. (b) atomically the core-shift in a plane perpendicular to the dislocation sense vector corresponds to removing or adding one atom. L goes along tilt axis [111]. P is the period vector. N is the boundary normal. (c) deleting two atoms in the left image shifts the boundary core to the configuration shown in the right image. (d) perspective view of core-shifted GB with core-shift depth (CSd) 9 atom layer deep and 18 deleted atoms (DA) in total, corresponding to CSd9DA18. (e) core-shifted GB [111] 13.17° step by step to build [112] core shifted emergent boundaries from CSd0DA0 to CSd12DA30. The top, middle and bottom panels are projected along the rotation axis L, the boundary normal N and the period vector P, respective. The cross-section is also shown after each three-view visualization. The atoms between two red arrows were deleted for the atomic coordinate relaxation and then the relaxation of simulation cell sizes. The atoms are selected out with the cohesive energy higher than $-3.466eV$, in contrast to the cohesive energy in bulk $\sim -3.54eV$, and hence only the defective atoms were shown. For CSd9DA18, a wedge corresponding to the sequence of deleted atoms is also shown.

**Figure 2 machine learning method**

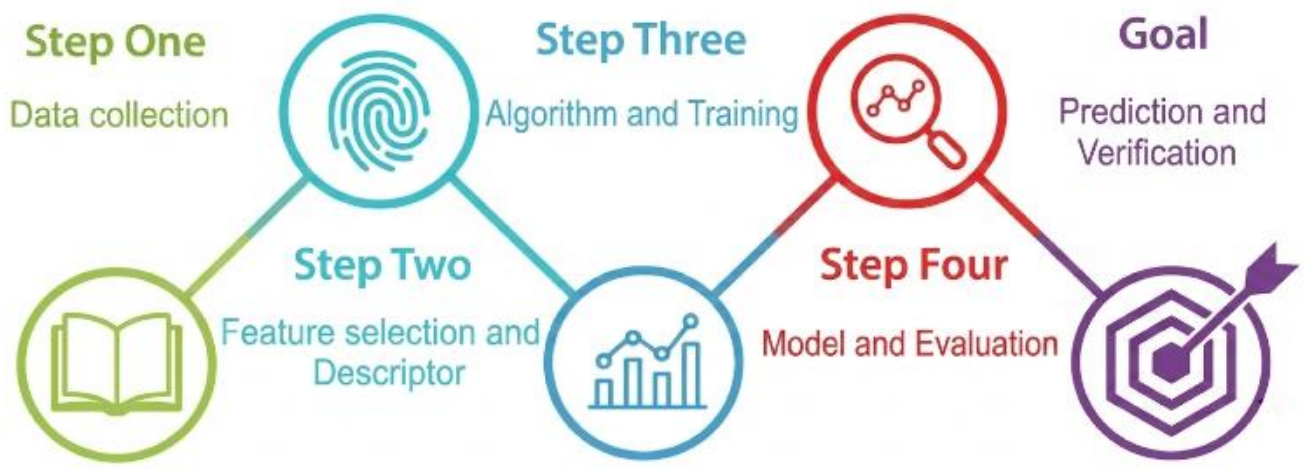

The method can be divided into two main steps. The first step includes data collection, feature selection, test of different algorithms, machine learning training and evaluation. The second one is to make prediction with the developed machine learning model.

**Figure 3 grain boundary structures at nanoscale**

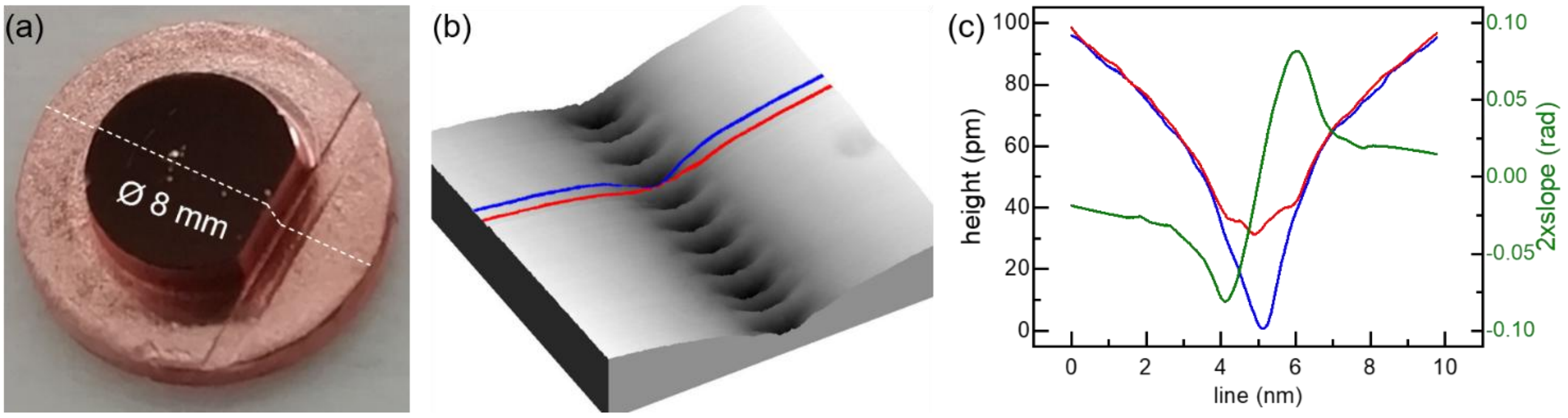

(a) macroscopic bicrystal of GB [111] $\mathbf{13.17}°$ with diameter 8 mm and thickness 2mm. The intersection between the grain boundary and the surface is shown schematically as a white line. (b) Perspective view of the surface across the grain boundary at nanometer scale from a 2D image of 10 nm × 10 nm. The imaging condition of scanning tunneling microscope is It = 20 pA and U = -0.01V. (c) topographic profile of two lines in (b) and two times of the local slope of the blue line.

**Figure 4 effect of the simulation supercell size and its relaxation on eGB energy**

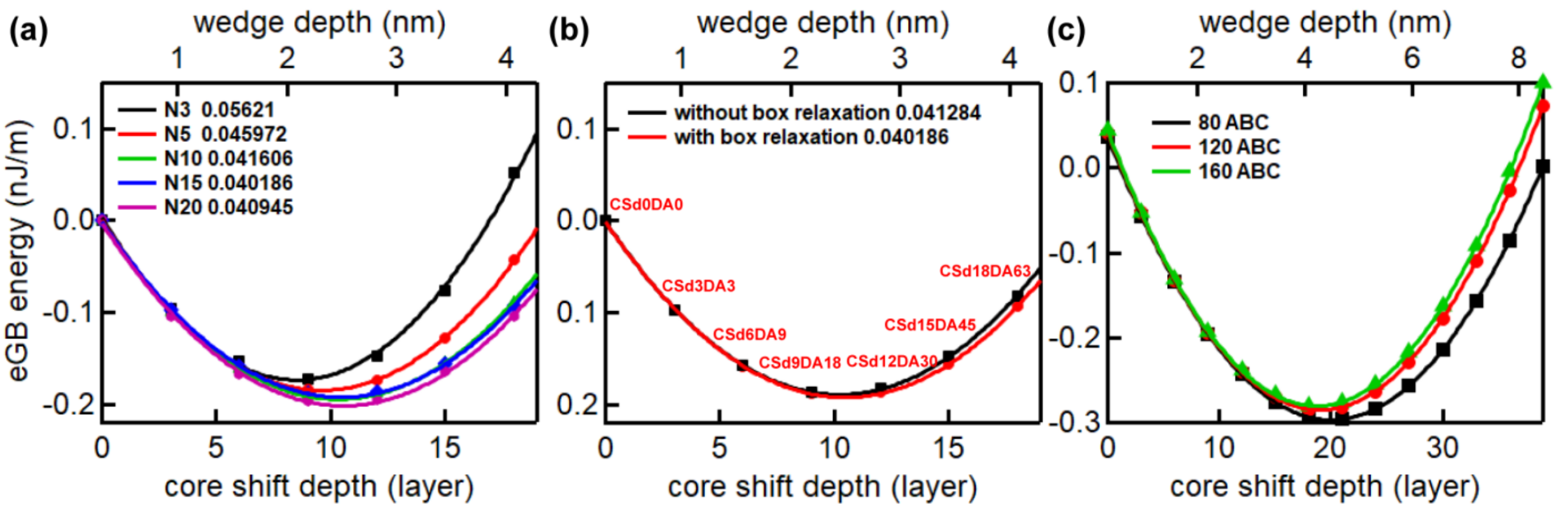

(a) the eGB energy for GB [111] $\mathbf{13.17}°$ as a function of core-shift depth $\boldsymbol{CSd}$ and the corresponding wedge depth, $\boldsymbol{h} = (\boldsymbol{CSd} + \mathbf{1.5}) \cdot \boldsymbol{a}/\sqrt{\mathbf{3}}$ (see the upper horizontal axis) for different simulation cell size along the boundary normal with the relaxation of the simulation cell length along the boundary normal. (b) the dependence of the eGB energy for GB [111] $\mathbf{13.17}°$ on the core-shift depth $\boldsymbol{CSd}$ and its corresponding wedge depth $\boldsymbol{h} = (\boldsymbol{CSd} + \mathbf{1.5}) \cdot \boldsymbol{a}/\sqrt{\mathbf{3}}$ (see the upper horizontal axis) with and without the relaxation along the boundary normal direction. (c) the dependence of the eGB energy for GB [111] $\mathbf{9.43°}$ on the wedge depth for three different film thickness, 80 ABC stacking unit, 120 ABC stacking unit and 160 ABC stacking unit.

**Figure 5 virial stress distribution of the core-shifted GB**

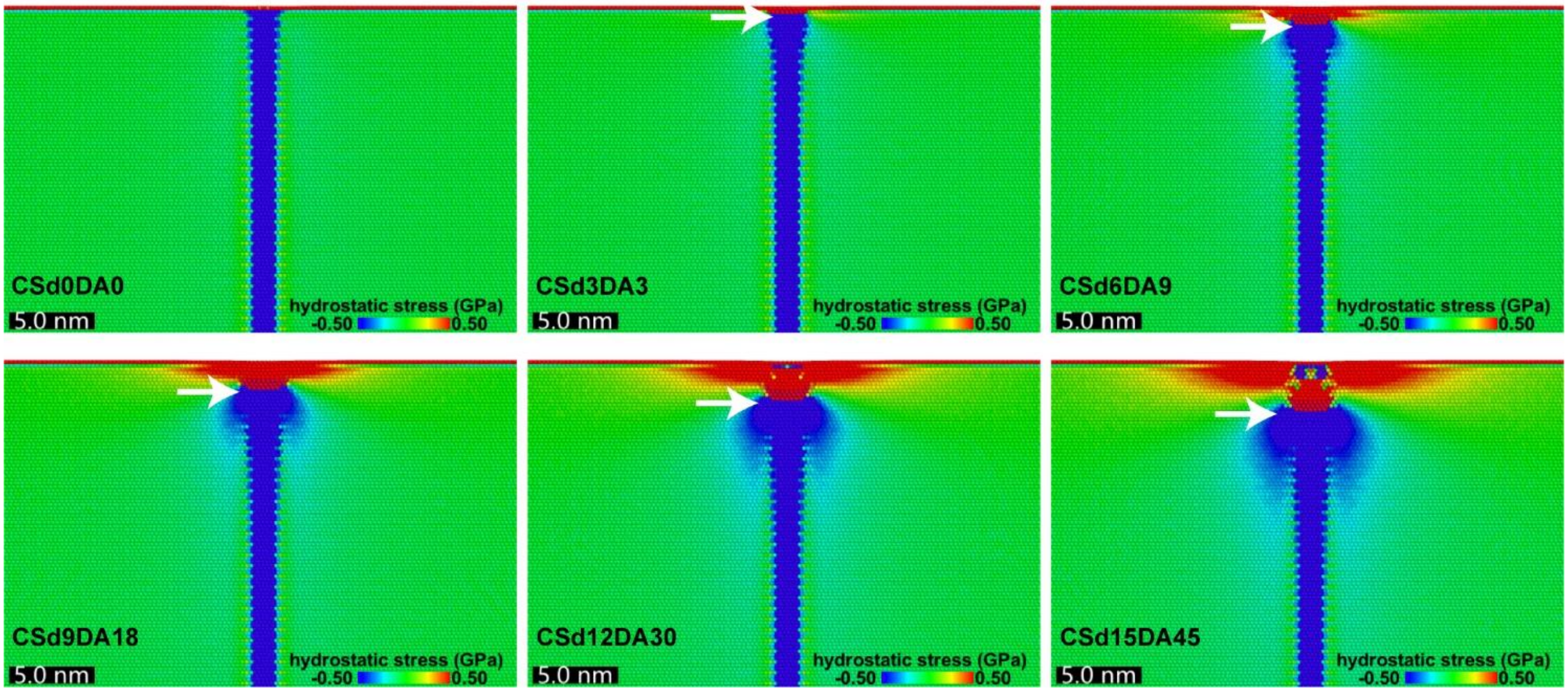

The atomic stress is calculated for the core-shifted GBs with the boundary core at the free surface along [112] direction and that deep in the bulk along [111] direction. The core-shift depth means the number of layers where atoms were deleted and the boundary cores were shifted toward [112]. (update the notation)

**Figure 6 build structures to delocalize the stress and CSd dependent eGB energy**

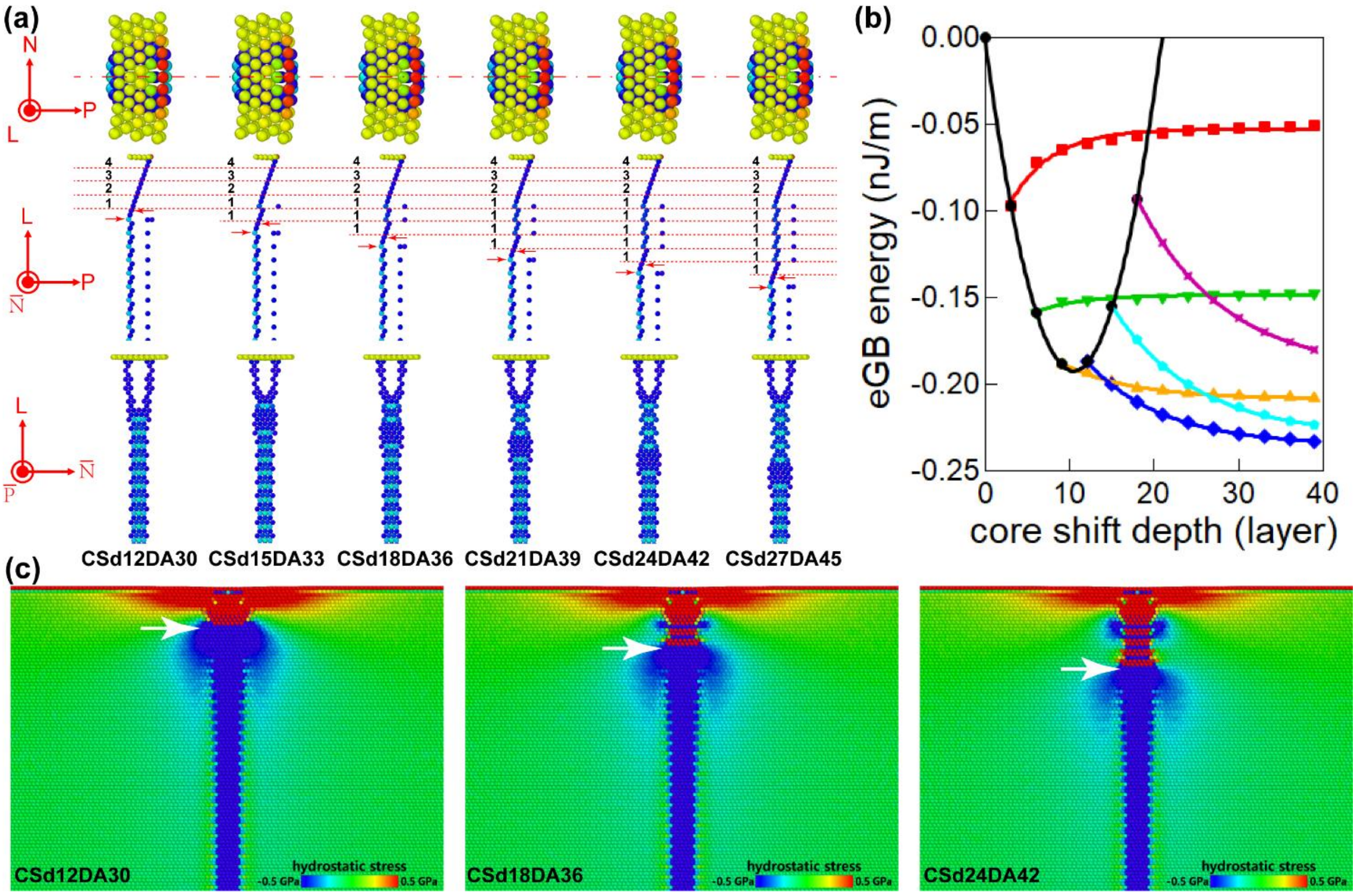

(a) the wedge tip blunting process with the core-shift method starting from CSd12DA30. The top, middle and the bottom panels are viewed along rotation axis, boundary normal, and period vector direction, respectively. The dot line in the middle panel lies just below the top 12 atomic layers. The rest eGB structures were obtained by core-shift further step by step through deleting the 3 atoms between two red arrows each time. Atoms are selected out by the cohesive energy greater than -3.47eV and hence only defect atoms are shown. The eGB energy for this list of structure was shown as the blue line in (b). The energy variation for other wedge tip blunting process beginning with corresponding [112] core-shifted eGBs, such as those shown in Fig. 1a, was also plotted out in (b). The energy for the [112] core-shifted eGBs were also drawn out as the red line. (c) the calculated atomic hydrostatic stress for three eGB structures, CSd12DA30, CSd18DA36, and CSd24DA42, which were shown in (a).

**Figure 7 STJ energies as a function of core-shift depth for eGB [111] $3.89°$ and eGB [111] $9.43°$**

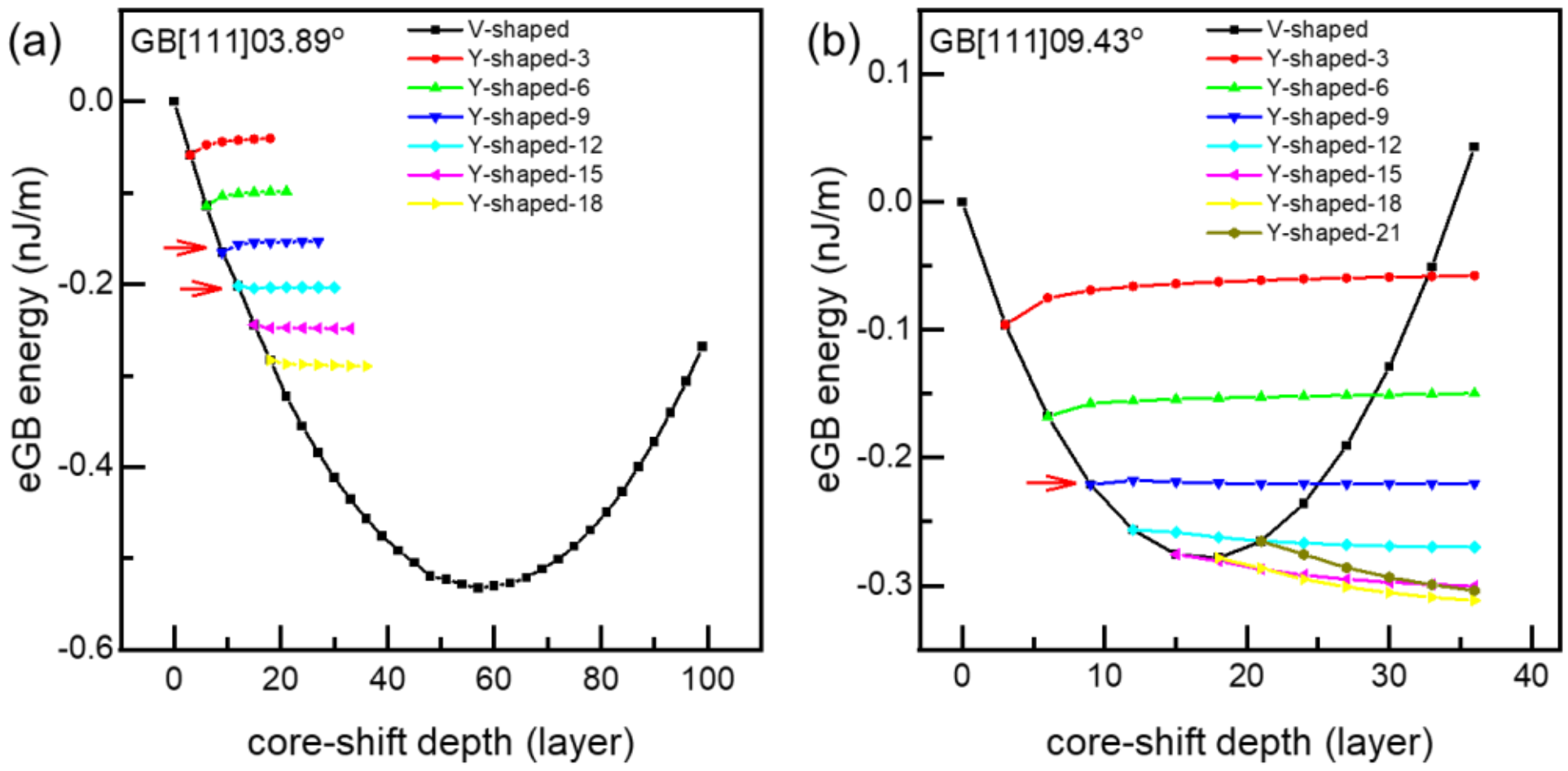

A plot of the eGB energy over core-shift depth. The points on the parabolic curve corresponds to [112] core shift grain boundaries. All other curves are fitted to the exponential formula. Each point on the exponential lines apart from the intersection points with the parabolic curve corresponds to one zipped

Y-shaped notch, built from the same zipped chevron V-shaped notch through blunting the wedge tip as shown in Fig. 6a.

**Figure 8 balance of the boundary energy and the elastic energy at a certain CSd**

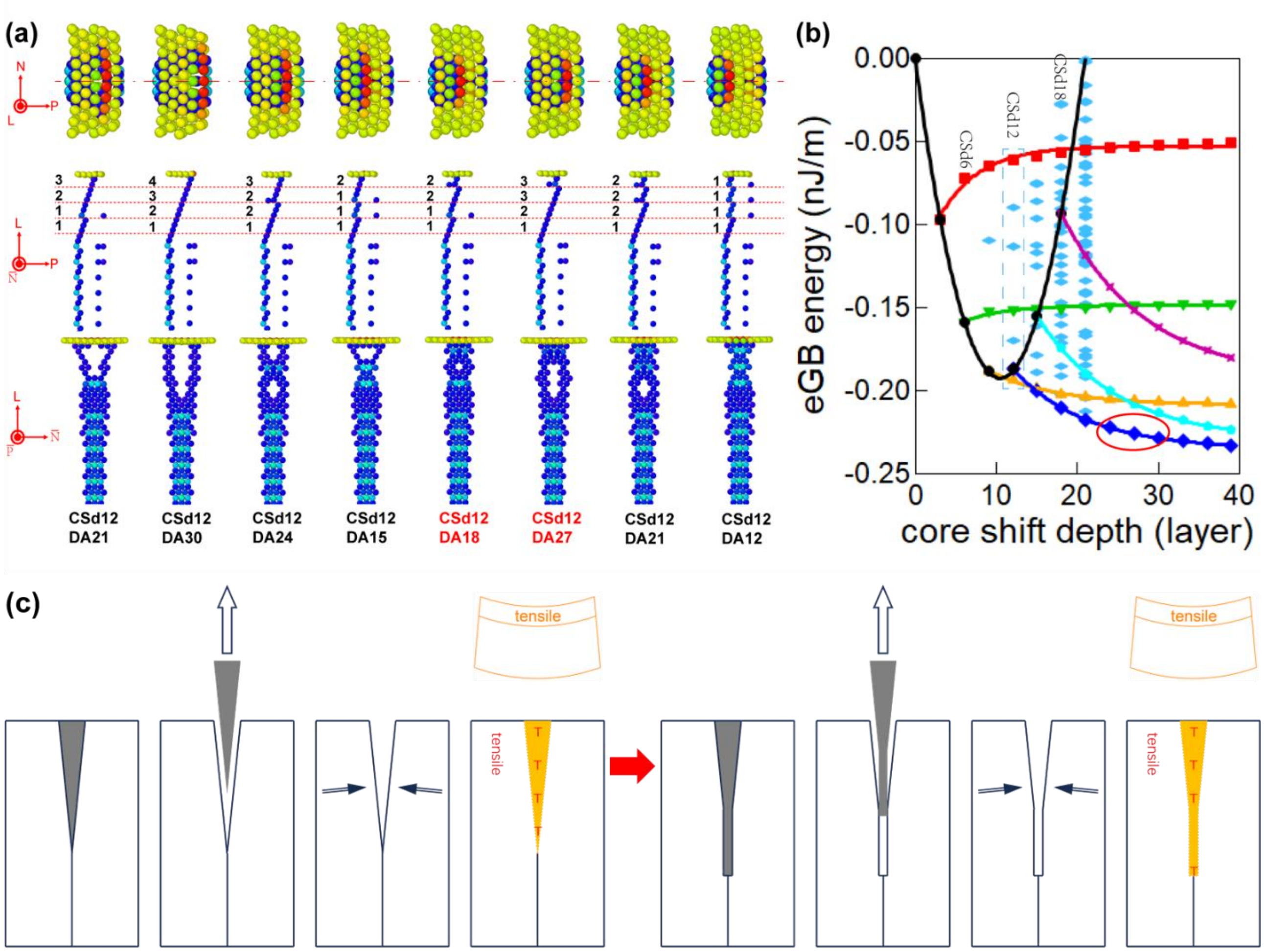

(a) a list of all the possible low energy eGB structures from the low energy to the high energy with the core-shift method at the fixed core shift depth. Note that there are two structures, CSd12DA18 and CSd12DA27, sharing almost the same eGB energy. The number of atoms deleted in each layer in different ABC stacking units were also labelled. Different ABC stacking units were separated with red dot lines. Atoms were selected out by the cohesive energy great than $\mathbf{-3.469}\boldsymbol{eV}$. (b) a plot of the eGB energy over core-shift depth. The [112] core-shift eGBs follow a parabolic curve in red. The energy values for 8 eGB structures with core-shift depth 12 shown in (a) are highlighted in the rectangle in (b). (c) the continuum model for the transition of the stable eGB structure from the zipped V-shaped notch to zipped Y-shaped notch. For both the zipped V-shaped notch and the zipped Y-shaped notch, the model includes four steps, these are initial structure, cut-out, zipping and final structure with local tensile stress. The substrate with films under tensile (compressive) residual stress tend to bend up (down).

**Figure 9 machine learning training and model**

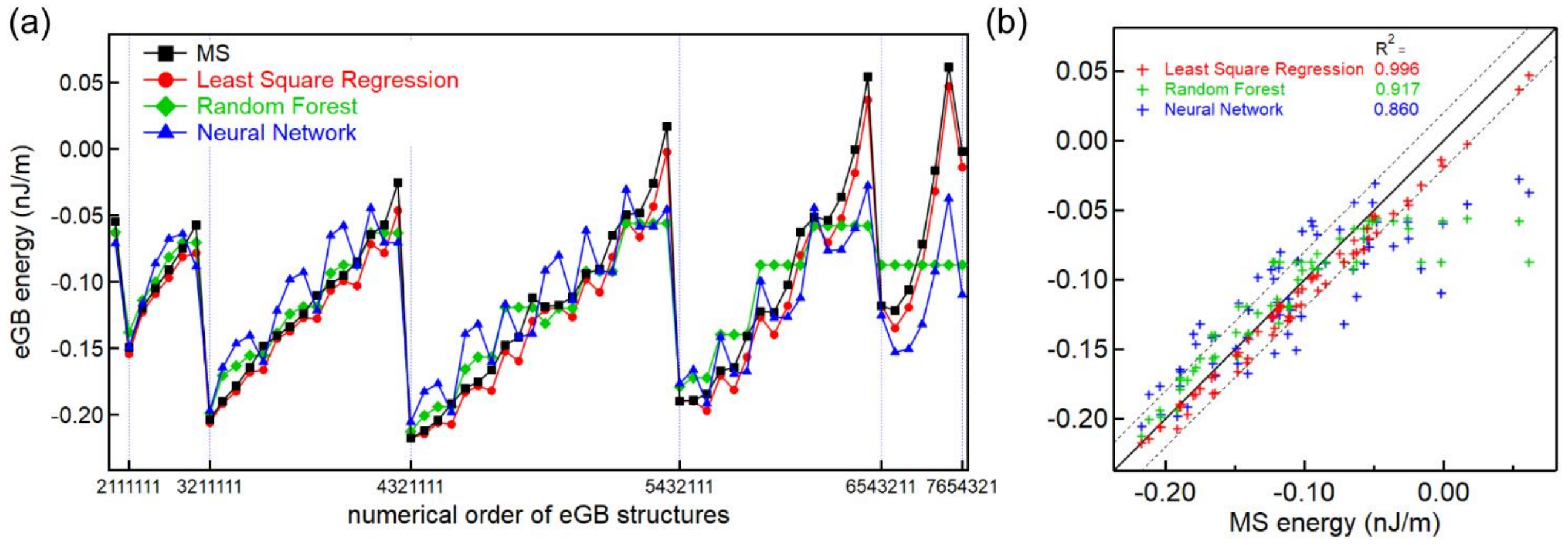

(a) Comparison of the eGB energy from machine learning with three different algorithms, that are neural network, random forest and LSR (degree=2), with the calculated eGB energy with MS for CSd=21. The machine learning model is obtained from the training. The training dataset includes all the data with CSd less than 21 and the additional 6 data points on the exponential lines with CSd=21. (b) eGB results predicted from ML model as a function of the results obtained from MS. The determination coefficient, R2 values, is for the testing data only. The solid line is the identity line and dashed lines show $\pm\mathbf{0.02}\,\boldsymbol{nJ/m}$ away from it.

**Figure 10 machine learning model and prediction**

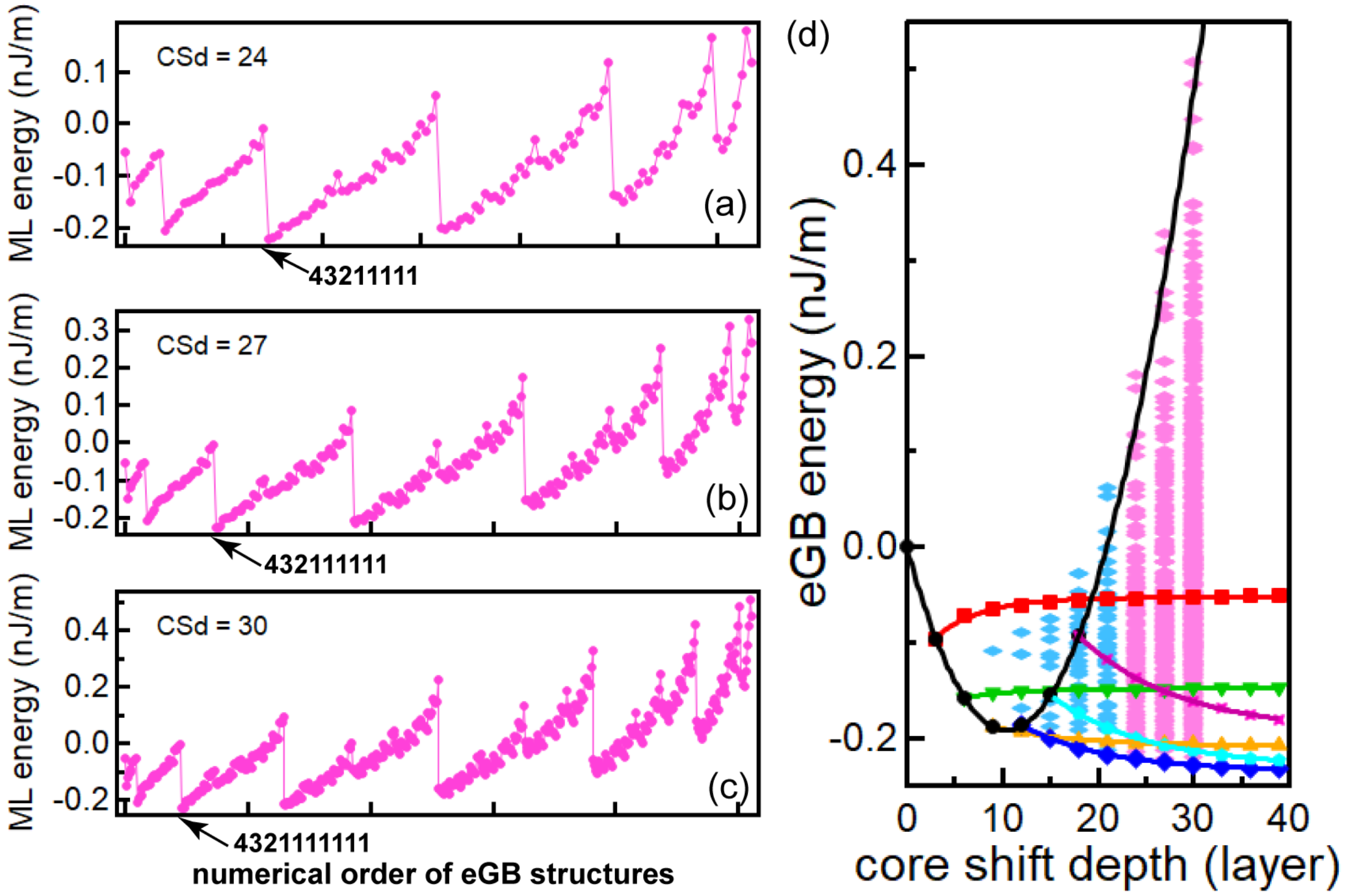

Prediction results with LSR (degree=2) algorithm for three different core shift depth (a) CSd=24, (b) CSd=27 and (c) CSd=30. The global minimum energies are indicated by arrows. The training dataset, from which the model is obtained, includes all the data with CSd less than 21 and the additional 6 data points on the exponential lines with CSd=24 for (a). Another 6 data points on the exponential lines with CSd=27 are added to the previous training dataset for (b). Another 12 data points on the exponential lines with CSd=27 and CSd=30 are added to the previous training dataset for (c). (d) eGB energy as a function of the core shift depth, with the predicted data included.

**Figure 11 STJ elastic energy calculated with MS and FEM method**

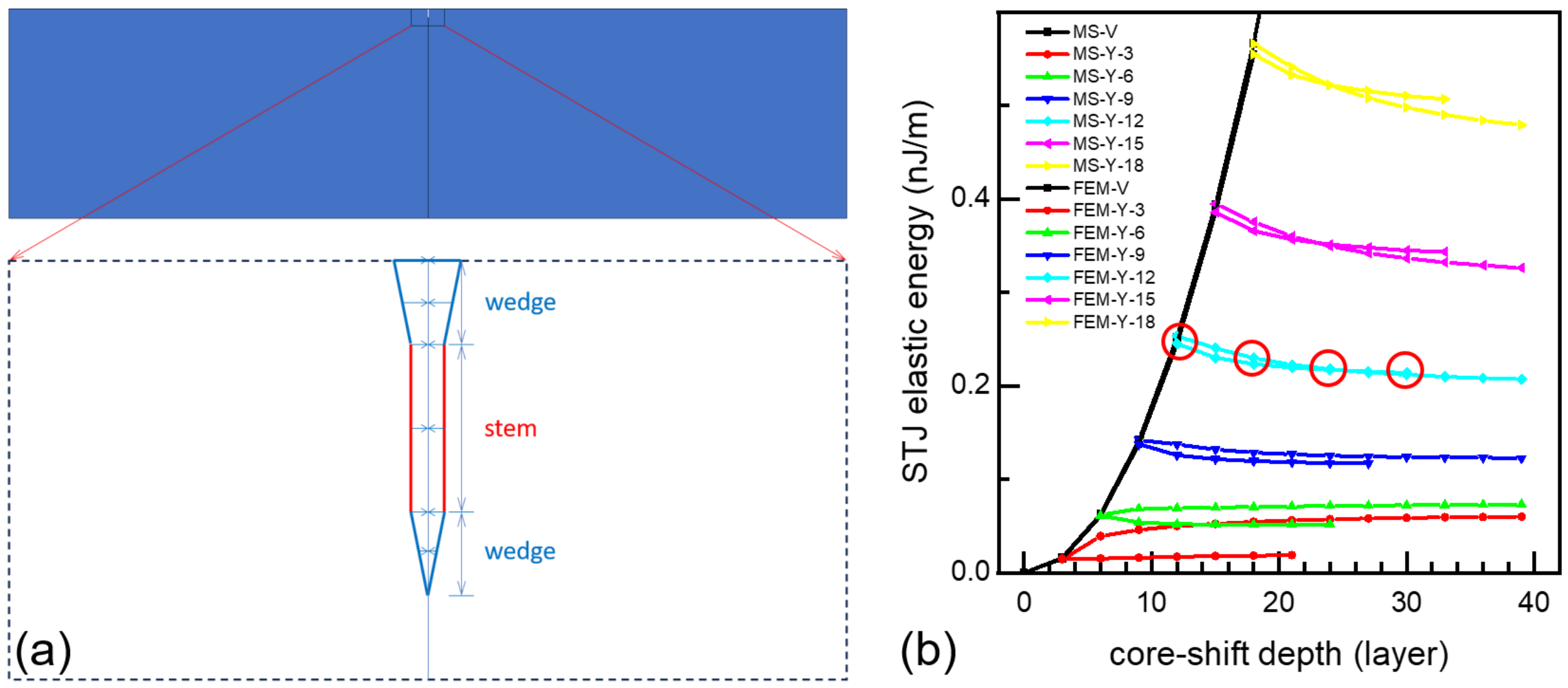

(a) the geometric configuration for FEM calculation. The displacement was used to build the initial configuration for relax. The top panel show the part and the lower panel show the close-up view of the STJ area. The Y-shapped notch includes wedge and stem. The Y-shaped notch turns into V-sahped notch in the case of no stem as shown in the black line. (b) Elastic energy as a function of the core shift depth calculated with MS and FEM method. The STJ elastic energies as a function of core-shift depth calculated with FEM overlaps with that calculated with MS. The first three points in circles correspond to three structures in Fig. 6c.